\documentclass{article}
\usepackage[utf8]{inputenc}
\usepackage{geometry}
\usepackage{amsmath}
\usepackage{amssymb}
\usepackage{amsfonts}
\usepackage{graphicx}
\usepackage{booktabs}
\usepackage{hyperref}
\usepackage{cite}
\usepackage{authblk}
\usepackage{algorithm}
\usepackage{algorithmic}

\title{\textbf{AntibodyDesignBFN: High-Fidelity Fixed-Backbone Antibody Design via Discrete Bayesian Flow Networks}}

\author[1]{Yue Hu}
\author[2]{Feng Tao}
\author[1]{Junqing Wang}
\author[3]{Yingchao Liu}

\affil[1]{School of Bioengineering, Qilu University of Technology (Shandong Academy of Sciences), No. 3501 Daxue Road, Jinan, Shandong, China}
\affil[2]{The Second Affiliated Hospital of Zhejiang University, School of Medicine, Zhejiang University, Hangzhou, 310053, China}
\affil[3]{Shandong Provincial Hospital, Shandong First Medical University}

\affil[ ]{Email: huyue@qlu.edu.cn, biogecko@126.com, wangjunqing@qlu.edu.cn, yingchaoliu@email.sdu.edu.cn}

\date{}

\begin{document}

\maketitle

\begin{abstract}
 The computational design of antibodies with high specificity and affinity is a cornerstone of modern therapeutic development. While deep generative models have demonstrated potential, they often struggle to balance high-fidelity geometric conditioning with the discrete nature of amino acid sequences. In this work, we present \textbf{AntibodyDesignBFN}, a novel framework for fixed-backbone antibody design based on \textbf{Discrete Bayesian Flow Networks (BFN)}. Unlike standard diffusion models, BFNs operate on a continuous probability simplex, enabling a fully differentiable generative process that seamlessly integrates geometric gradients. By combining a lightweight \textbf{Geometric Transformer} with \textbf{Invariant Point Attention (IPA)} and a resource-efficient training strategy, our model establishes a new state-of-the-art. Evaluations on a rigorous \textbf{2025 temporal test set (43 complexes)} demonstrate that AntibodyDesignBFN achieves an unprecedented \textbf{Amino Acid Recovery (AAR) of 67.8\%}, significantly outperforming leading graph-based baselines. Furthermore, the model is highly efficient, enabling millisecond-scale inference on consumer-grade hardware. AntibodyDesignBFN thus offers a powerful, accessible, and mathematically robust framework for next-generation antibody engineering. Code and model checkpoints are available at \url{https://github.com/YueHuLab/AntibodyDesignBFN} and \url{https://huggingface.co/YueHuLab/AntibodyDesignBFN}.
\end{abstract}

\section{Introduction}

Antibodies are critical components of the adaptive immune system and represent a dominant class of biotherapeutics. Their binding specificity is primarily governed by six complementarity-determining regions (CDRs) in conventional antibodies, or three CDRs on the heavy chain in the case of single-domain antibodies (nanobodies or VHHs). Among these, the third CDR of the heavy chain (H-CDR3) is typically the most diverse and functionally significant loop for antigen recognition. The "inverse folding" problem in antibody engineering often entails designing specific amino acid sequences—particularly for the variable CDR loops—that adopt a desired 3D backbone conformation and bind a target epitope, while keeping the rest of the scaffold fixed.

Traditional approaches often rely on physics-based energy minimization (e.g., Rosetta \cite{rosetta}), which is computationally intensive and limited by the accuracy of energy functions. In recent years, deep learning methods have introduced generative paradigms, ranging from autoregressive models (e.g., ProteinMPNN \cite{proteinmpnn}) to joint sequence-structure diffusion models like DiffAb \cite{luo2022antigen} and AlphaPanda \cite{hu2025alphapanda}.

The emergence of \textbf{Bayesian Flow Networks (BFNs)} \cite{graves2023bayesian} has provided a powerful new alternative to diffusion models. BFNs reformulate generation as a Bayesian inference process, where the model iteratively updates its belief about the data distribution. \textbf{IgCraft} \cite{igcraft} and \textbf{AbBFN} \cite{abbfn} have recently demonstrated the power of BFNs for paired antibody sequence generation and motif scaffolding. However, these models primarily focus on sequence-space evolution or involve joint generation of multiple modalities including metadata, often without explicit structural conditioning on the antigen. Our work, \textbf{AntibodyDesignBFN}, extends this paradigm by conditioning the discrete BFN generative process on a full 3D atomic backbone. Crucially, this allows our model to explicitly "see" the antigen epitope geometry during generation, ensuring that the designed sequences are not only naturally diverse but also structurally and electrostatically compatible with the target interface.

\section{Methodology}

\subsection{Discrete Bayesian Flow Networks}

We formulate the antibody sequence design problem using the \textbf{Discrete Bayesian Flow Network} framework. Let $\mathbf{x}$ be a ground-truth amino acid sequence of length $L$. In our fixed-backbone setting, the length $L$ is strictly determined by the input 3D structure. The Bayesian distribution is defined over the logits $\mathbf{\theta} \in \mathbb{R}^{L \times K}$ of the categorical distribution for the entire sequence simultaneously. This means the model updates its belief about every residue in parallel, capturing global dependencies without autoregressive bias.

The core distinction of BFNs is that they model the \textbf{parameters} (logits) of the sequence distribution rather than the sequence tokens themselves. This allows us to use continuous Gaussian processes to generate discrete data.

\subsubsection{The Bayesian Sender (Forward Process)}
The "Sender" distribution $P_S(\mathbf{\theta} | \mathbf{x}; t)$ defines how our information (belief) about the sequence evolves over continuous time $t \in [0, 1]$. This is fundamentally different from Diffusion Models:
\begin{itemize}
    \item \textbf{Diffusion Models}: Corrupt the data $\mathbf{x}$ directly (e.g., by flipping tokens or adding Gaussian noise to embeddings).
    \item \textbf{BFNs}: Transmit the true data $\mathbf{x}$ through a noisy channel into the parameter space $\mathbf{\theta}$.
\end{itemize}

The parameters $\mathbf{\theta}(t)$ are sampled from a Gaussian distribution:
\begin{equation}
    \mathbf{\theta}(t) \sim \mathcal{N}(\beta t \cdot \mathbf{e}_{\mathbf{x}}, \beta t \cdot \mathbf{I})
\end{equation}
Here, $\beta$ is a signal-to-noise ratio parameter. At $t=0$ (Prior), the distribution $\text{Softmax}(\mathbf{\theta})$ is Uniform (Maximum Uncertainty). At $t=1$ (Data), the mean is $\beta \mathbf{e}_{\mathbf{x}}$, and the distribution is sharp (Maximum Certainty).

\subsubsection{The Geometric Receiver (Architecture)}
The "Receiver" is a neural network $\Psi$ that estimates the true data distribution given the noisy parameters $\mathbf{\theta}$ and time $t$. The input to the network is the current "belief" about the sequence, represented as $\mathbf{p}_{in} = \text{Softmax}(\mathbf{\theta})$. This soft probability distribution is concatenated with sinusoidal time embeddings and passed to the geometric encoder:
\begin{equation}
    \mathbf{p}_{out} = \Psi(\mathbf{p}_{in}, t, \mathcal{G}_{backbone})
\end{equation}
We employ a \textbf{Geometric Transformer} utilizing \textbf{Invariant Point Attention (IPA)} layers. This explicitly conditions the sequence belief update on the 3D coordinates of the N, $C_\alpha$, C atoms. The attention mechanism calculates weights based on both the semantic similarity of the intermediate features and the Euclidean distance between residues in the rigid backbone frame. This ensures that residues spatially close in the 3D structure can communicate efficiently, even if they are distant in the primary sequence.

\subsubsection{Continuous-Time Loss Function}
The training objective simplifies elegantly to a weighted reconstruction loss:
\begin{equation}
    \mathcal{L}(\mathbf{x}) = \beta \int_0^1 \mathbb{E}_{\mathbf{\theta} \sim P_S(\cdot|\mathbf{x}; t)} { || \mathbf{e}_{\mathbf{x}} - \hat{\mathbf{p}}_{out}(\mathbf{x} | \mathbf{\theta}, t) ||^2 } dt
\end{equation}
In practice, we minimize a cross-entropy loss $-\log(\mathbf{p}_{out}(\mathbf{x}_{true}))$, which is numerically more stable and optimizes the same objective.

\subsection{Bayesian Sampling (Inference)}
The generation process starts with a uniform prior $\mathbf{\theta}_0 = \mathbf{0}$. At each step $k$ (of $N$ total steps), we perform a Bayesian update:
\begin{equation}
    \mathbf{\theta}_{k+1} = \underbrace{\mathbf{\theta}_k}_{ \text{Prior}} + \underbrace{\frac{\beta}{N} \hat{\mathbf{p}}_{out}(\mathbf{\theta}_k, t_k)}_{ \text{Evidence}} + \underbrace{\sqrt{\frac{\beta}{N}} \mathbf{\epsilon}_k}_{ \text{Uncertainty}}
\end{equation}
This process can be viewed as an iterative refinement where the sequence emerges from the noise, guided by the structural constraints.

\subsection{Training Strategy}
We utilized gradient accumulation (\texttt{accum\_steps=8}) to overcome limits and simulate a larger effective batch size, leveraging an \textbf{Apple M3 Studio with 256GB unified memory} for high-throughput training. We employed FP16 mixed-precision training (via \texttt{torch.cuda.amp} or \texttt{torch.mps}) and a linear warmup strategy for the first 1,000 steps followed by a constant learning rate of $1e-4$.
\textbf{Remarkably, the model is highly efficient: we verified that both training and fast inference can be successfully performed on consumer-grade hardware, such as a Mac mini M4 (16GB), demonstrating broad accessibility.}

Furthermore, we prioritized \textbf{data quality over quantity} during training set construction. Inspired by recent advances in antibody-specific modeling \cite{abmpnn, antibmpnn}, we rigorously filtered the SAbDab dataset to exclude low-resolution structures and entries with missing residues in CDR regions, ensuring the model learns from high-fidelity geometric signals.

\section{Experiments}

\subsection{Experimental Setup}
\textbf{Dataset}: We utilized the Structural Antibody Database (SAbDab), filtered for non-redundancy (95\% sequence identity clusters) with a training cutoff date of 2022.

\textbf{Test Set}: To rigorously evaluate the generalization capability of AntibodyDesignBFN and prevent any potential sidechain information leakage (a common pitfall in fixed-backbone design), we curated a novel test set consisting of \textbf{43 antibody-antigen complexes released in 2025}, which were strictly excluded from our training data. Crucially, we pre-processed these structures by mutating all residues within the CDR loops to \textbf{Poly-Alanine (PolyA)}, removing all native sidechain information. This forces the model to recover the sequence solely from the backbone geometry and the antigen context. All structures contain both the antibody Fv fragment and a bound antigen, ensuring that the fixed-backbone condition reflects real binding geometries. The PDB IDs are: 8S6T, 8S6V, 8UHP, 8UIG, 8UIH, 8UJI, 8V13, 8YYZ, 8Z2V, 8Z39, 8Z3A, 9C7X, 9CDS, 9CFD, 9CQA, 9CR9, 9D41, 9DBO, 9DQ3, 9DQ4, 9DST, 9EHT, 9EZE, 9FIK, 9GGP, 9GOX, 9H4R, 9J8A, 9JCY, 9JD1, 9M5B, 9MIC, 9MID, 9MIF, 9NJY, 9NPI, 9OAR, 9P4C, 9PI9, 9PRU, 9PWN, 9RM2, 9UK5.

\subsection{Results}

We evaluate performance using Amino Acid Recovery (AAR). To provide a rigorous baseline, we also evaluated \textbf{ProteinMPNN} \cite{proteinmpnn} on the same 2025 dataset. Table \ref{tab:results} presents the comparative results using the finetuned model.

\begin{table}[h]
\centering
\caption{Comparative Performance on the 2025 Temporal Test Set (43 Antibody-Antigen Complexes). The reported AAR values reflect the robustness of sequence recovery under fixed-backbone conditions.}
\label{tab:results}
\begin{tabular}{l|cccccc|c}
\toprule
\textbf{Model} & \textbf{H1} & \textbf{H2} & \textbf{H3} & \textbf{L1} & \textbf{L2} & \textbf{L3} & \textbf{Avg} \\
\midrule
\textbf{AntibodyDesignBFN} & \textbf{74.8} & 55.9 & 48.4 & \textbf{79.6} & \textbf{82.1} & \textbf{66.0} & \textbf{67.8} \\
ABMPNN & 64.7 & 55.2 & \textbf{49.3} & 63.0 & 57.5 & 55.9 & 57.6 \\
AntiBMPNN (000) & 59.7 & \textbf{56.8} & 39.7 & 60.3 & 52.5 & 53.4 & 53.7 \\
Vanilla MPNN & 42.2 & 44.0 & 35.0 & 42.3 & 45.8 & 41.4 & 41.8 \\
\bottomrule
\end{tabular}
\end{table}

\section{Discussion}

\subsection{Performance Analysis: Linking AAR to Structural Complexity}
AntibodyDesignBFN achieves an unprecedented Amino Acid Recovery (AAR) of \textbf{67.8\%}, with performance variance across regions strongly correlated with structural complexity. We observe peak performance in canonical loops like L2 (82.1\%), L1 (79.6\%), and H1 (74.8\%), where the model effectively "locks in" sequences for these rigid restricted scaffolds. In contrast, the H-CDR3 loop presents the greatest challenge due to its extreme length diversity and conformational flexibility. Our AAR of 48.4\% is comparable to the graph-based ABMPNN (49.3\%). This reflects the intrinsic multimodal nature of H3 design, where many distinct sequences can fold into similar backbone geometries. BFN's continuous probabilistic modeling allows it to explore this complex manifold rather than collapsing to a single mode.

\subsection{Why BFN Succeeds: Differentiable Optimization and Continuous Flow}
The AntibodyDesignBFN framework offers unique advantages that stem from its continuous mathematical formulation:
\begin{itemize}
    \item \textbf{Sequence Gradient Optimization}: A critical advantage of our approach is that the entire generative process is fully differentiable. Unlike discrete autoregressive or diffusion models, BFNs operate on the continuous probability simplex. This allows gradients to flow from any downstream objective directly back to the sequence parameters (logits). This capability is pivotal for potential future applications where we optimize sequences not just for recovery, but for specific binding affinities or developability properties.
    \item \textbf{Continuous-Time Manifold}: The model treats amino acid identity not as a discrete token flip but as a continuous evolution of a belief distribution. This smoothness helps traverse the complex energy landscapes of variable loops like H3 without getting trapped in local optima, a common failure mode of discrete corruption processes.
    \item \textbf{High-Fidelity Geometric Conditioning}: By integrating Invariant Point Attention (IPA) \cite{jumper2021highly}, our model effectively captures the subtle orientational constraints of sidechains within the CDR loops.
\end{itemize}

\subsection{Efficiency and Accessibility}
A major feature of AntibodyDesignBFN is its minimal hardware footprint. While achieving SOTA accuracy, the model remains lightweight due to the efficient BFN parametrization. \textbf{Inference is particularly fast}, taking only milliseconds. We expressly verified deployment on consumer-grade hardware, including the \textbf{Mac mini M4 (16GB)}, ensuring this tool is accessible to labs without cluster infrastructure.

\subsection{Data Strategy: Quality over Quantity}
Following the pioneering data-centric insights demonstrated by \textbf{AbMPNN} \cite{abmpnn} and \textbf{AntiBMPNN} \cite{antibmpnn}, we prioritized the quality of geometry over the quantity of sequences. While the OAS database \cite{oas} offers billions of sequences, it lacks structural ground truth. Therefore, we trained exclusively on the \textbf{SAbDab} database \cite{sabdab}, applying a rigorous resolution cutoff of \textbf{5.0 \AA}. This strategy ensures that our model learns from experimentally determined physical constraints (via X-ray or Cryo-EM) rather than noisy predicted structures. Our results confirm that for the inverse folding task, high-fidelity structural labels are far more valuable than massive unbound sequence data.

\section*{AI Usage Declaration}
This manuscript was prepared with the assistance of large language models, specifically \textbf{Gemini 2.0} and \textbf{Google Antigravity}, which were used for drafting text, refining mathematical formulations, checking LaTeX syntax, and exploring related literature. All content, data, and conclusions have been manually reviewed, verified, and authenticated by the authors, who take full responsibility for the accuracy, integrity, and originality of the work.

\bibliographystyle{unsrt}
\bibliography{references}

\end{document}